# Neutral particle Mass Spectrometry with Nanomechanical Systems

Eric Sage[1,2], Ariel Brenac[3,4], Thomas Alava[1,2], Robert Morel[3,4], Cécilia Dupré[1,2], Mehmet Selim Hanay[5†], Michael L. Roukes[5], Laurent Duraffourg[1,2], Christophe Masselon[6,7,8], Sébastien Hentz[1,2,*]

1. Univ. Grenoble Alpes, F-38000 Grenoble, France
2. CEA, LETI, Minatec Campus, F-38054 Grenoble, France
3. Univ. Grenoble Alpes, INAC-SP2M, F-38000 Grenoble, France
4. CEA, INAC- SP2M, F-38000 Grenoble, France
5. Kavli Nanoscience Institute and Departments of Physics, Applied Physics, and Bioengineering, California Institute of Technology, MC 149-33, Pasadena, California 91125 USA
6. CEA, IRTSV, Biologie à Grande Echelle, F-38054 Grenoble, France
7. INSERM, U1038, F-38054 Grenoble, France
8. Université Joseph Fourier, Grenoble 1, F-38000, France
  † Present address: Department of Mechanical Engineering and UNAM, Bilkent University, Ankara 06800, Turkey
  * sebastien.hentz@cea.fr

**Current approaches to Mass Spectrometry (MS) require ionization of the analytes of interest. For high-mass species, the resulting charge state distribution can be complex and difficult to interpret correctly. In this article, using a setup comprising both conventional time-of-flight MS (TOF-MS) and Nano-Electro-Mechanical-Systems-based MS (NEMS-MS)** *in situ,* **we show directly that NEMS-MS analysis is insensitive to charge state: the spectrum consists of a single peak whatever the species' charge state, making it significantly clearer than existing MS analysis. In subsequent tests, all charged particles are electrostatically removed from the beam, and unlike TOF-MS, NEMS-MS can still measure masses. This demonstrates the possibility to measure mass spectra for neutral particles. Thus, it is possible to envisage MS-based studies of analytes that are incompatible with current ionization techniques and the way is now open for the development of cutting edge system architectures with unique analytical capability.**

Of all analytical techniques, Mass Spectrometry has grown fastest over the past two decades[1,2]; it is now recognized as an essential tool in a variety of fields of modern research ranging from environmental chemistry[3] to structural biology[4], or even astrophysics[5]. After ionization of the analytes of interest, the spectrum generated is interpreted on the mass-to-charge ratios of the different ions. Thanks to the ground-breaking development of soft ionization methods[6,7], MS is no longer limited to applications involving small volatile molecules, but can be used to study massive, supramolecular assemblies[8–10]. Nevertheless, routine use of MS in the MDa to GDa range remains challenging as, in this mass range, correct analysis requires a large number of charges per particle to produce mass-to-charge ratios compatible with the detection capacities of most mass analyzers[11]. The enormous number of charge states obtained may produce difficult to interpret spectra due to overlapping peaks[9] as well as drastic changes of physical state compared to buffered solution. In some cases, the study of molecules in their "native" form (due to, for example, dissociation[9,12]) can be made complex. Moreover, it has been estimated that only one analyte ion out of $10^3$-$10^5$ generated at ambient pressure is generally detected[13]. This state of affairs has led to extensive efforts to understand and improve the efficiency of ionization and ion-transmission[13,14] as well as ion detection, in particular for high masses[8,15].

Mass sensing has been widely performed since the mid-nineties using Micro- and Nanomechanical resonators which detect changes in resonance frequency upon mass accretion on their surface[16–18]. A mass resolution of a few Da was recently reported[19]. While the first NEMS mass spectrometer was being developed[20], a procedure to simultaneously deduce both the mass and position of a single particle present on the resonator by monitoring several resonance modes was designed by the same group and later performed at the microscale[21]. This effort eventually culminated in real-time acquisition of NEMS-MS mass spectra for single particles[22]. In systems equipped with standard ion sources, this experiment demonstrated that NEMS can detect and measure ionized molecules. However, NEMS-based measurements have never been benchmarked against the results obtained with a well-established MS system. This lack of comparison probably stems from the fact that NEMS operate in a mass range where there is no accepted mass standard (i.e., the atomic or molecular calibrants used in MS[23,24]). Furthermore, NEMS-MS has an attribute distinguishing it from all other MS systems – it can measure **both** neutral and ionized molecules – but this attribute has yet to be exploited. In this article, we demonstrate the unique features of NEMS-MS using an experimental setup where both neutral and ionized metallic nanoclusters can be produced on demand. This novel hybrid setup makes it possible to acquire mass spectra for these nanoclusters by both NEMS-MS and TOF-MS.

## Results

**Setup**
The setup consisted of four main vacuum chambers (see Figure 1a): a nanocluster source, an intermediate chamber equipped with electrostatic deflection plates, a deposition chamber and an in-line TOF mass spectrometer. Metallic nanoclusters were generated using a sputtering gas aggregation technique[25]. Nanoclusters were then expelled into the vacuum deposition chamber ($10^{-5}$ mbar) through a differential pumping stage. The metallic cluster sizes and the cluster deposition rate could be tuned by adjusting the operating parameters (see Supplementary Methods). The deposition rate was measured thanks to a Quartz Crystal Microbalance (QCM) placed on a translational stage. Upon retraction of this stage, the nanomechanical resonator shown in Figure 1b was exposed to the cluster flux. When both NEMS and QCM were retracted, the particle flux could enter the acceleration region of the in-line TOF mass spectrometer, where the mass-to-charge distribution of charged particles was measured. Transmission Electronic Microscopy (TEM) performed on nanoclusters deposited on TEM grids showed that the TOF module accurately detects the total mass distribution for the clusters[26] (see Supplementary Methods). The configuration of the deposition chamber makes it possible to perform QCM, TOF-MS and NEMS-MS measurements sequentially; this setup ensures that the cluster distribution remains unchanged within the timeframe of the experiment (typically from 30 min to several hours) (detailed in Supplementary Methods).

**Comparison of TOF-MS and NEMS-MS**
For our experiments, we chose tantalum as the analyte as it is both dense (16.6 g.cm$^{-3}$) and readily condenses into large clusters. The cluster populations generated follow a log normal distribution law with mean masses of up to 4.5 MDa (diameter up to 9.5 nm). The TOF and NEMS-MS mass spectra acquired for various populations, with mean masses down to 470 kDa, were compared. Excellent matching of both the central mass and the width of the main peak distribution was obtained between the two techniques, particularly for larger particles (see Supplementary Figure 5). As the mass resolution of a nanomechanical sensor is constant over the whole mass range[27] it provides lower relative uncertainty for heavier particles. This contrasts with current MS systems where the resolving power (capacity to distinguish between two peaks with similar masses) decreases as mass increases[11]. A comparison between NEMS-MS and TOF-MS for a cluster population centered on 2420 kDa is presented in Figure 2. NEMS-MS spectra displayed a single peak, but additional peaks were present on the TOF-MS spectrum. These additional peaks were visible in a wide range of operating conditions for cluster generation (clusters typically measured 4 nm or more). Regardless of the mass distribution generated, the mass ratios between these peaks remained constant and equal to $\frac{1}{2}$ and $\frac{1}{3}$. However, multimodal size distributions were never observed with TEM measurements[26], indicating that the true particle mass distribution is represented by a single peak. The multiple peaks detected with TOF-MS can thus be attributed to the presence of multiple charge states in the compound analyzed, which is a common phenomenon in current MS. Thus, different charge states lead to several peaks because ion-based MS is sensitive to the mass-over-charge ratio $\left(\frac{m}{z}\right)$. This phenomenon can become problematic when a number of charge states co-exist, since it decreases the signal-to-noise ratio (because the ions of interest are spread over multiple peaks). When this occurs with mixtures to be analyzed, spectral interpretation requires specific procedures to deal with overlapping multiply-charged envelopes[9,28,29]. Moreover, the interpretation of spectra containing multiple charge states with low isotopic resolution becomes a real issue, since it can be difficult to distinguish between the presence of species at a fraction of the mass

$\left(\frac{\frac{1}{x}m}{z}\right)$ of a particular analyte of mass *m* and the presence of a multiply-charged version of the same analyte $\left(\frac{m}{xz}\right)$[30]. To identify a species, the instrument resolution must be sufficient to resolve multiple peaks, and assign charge numbers to deduce a precise mass. NEMS-MS detection provides a clear, single peak, unlike m/z detection which displays multiple charge states. Indeed, the resolution - 70 kDa as directly measured based on Allan deviation[31] - is amply sufficient to resolve any of the apparent peaks present on the TOF-MS spectrum, and indeed, in other acquisitions, clusters with mean masses below 500 kDa were successfully measured. Figure 2 clearly shows that NEMS-MS spectra provide a clean signal, untainted by the compounding effect of multiple charge states. Consequently, this spectrum unequivocally shows that the nanoparticle population produced can be represented by a single distribution with a well-defined peak value corresponding to its mass. NEMS-MS can thus avoid possible spectrum misinterpretation or loss of information caused by peak overlap. This attribute could be of great interest when analyzing mixtures, i.e., for the vast majority of MS measurements.

**Mass spectrum of neutral particles**
The results presented here demonstrate that NEMS-MS is insensitive to charge and can be used to directly measure mass. This potentially represents a major paradigm shift in MS since ionization could be circumvented altogether. Without ionization, it will be possible to analyze compounds in their native state, regardless of their charge. To confirm that this is possible, we went on to investigate whether the NEMS device could detect neutral particles. A DC voltage of 40 V was applied to the deflection plates located in the intermediate chamber. This voltage caused charged particles (known to form in the cluster source) to deviate from the straight flow line and collide with the plate (producing a black spot on the plates, see Figure S2d) while neutral particles were unperturbed and continued on toward the deposition chamber. TOF and NEMS-MS spectra were acquired with and without electrostatic deflection: Figure 3 shows the first mass spectra obtained for neutral particles. While no signal was observed with the TOF module (confirming the absence of charged particles), NEMS-MS successfully measured the mass distribution for the neutral particle population. Once again, the peak mass of the undeflected singly-charged TOF spectra correlated very well with both the deflected (neutrals only) and undeflected (both ions and neutrals) NEMS-MS spectra, confirming that the mass distributions for neutral and ionized clusters were identical.

**Discussion**
Analysis of large (supra-) molecular entities is extremely challenging using ion-based MS methods as a high number of charges is required to produce mass-to-charge ratios compatible with the detection capacity of most mass analyzers[11]. Very few instruments are capable of resolving successive charge states in this high mass/ high charge regime[10]. Moreover, not all species ionize with equal efficiency[24,30]. Thus, MS is inherently non-quantitative and internal standards must be used to calibrate the analytical response[23]. Although Charge Detection MS circumvents charge state assignment issues by simultaneously measuring the charge and the m/z of individual ions, it still requires highly charged particles (hundreds to thousands) when analyzing massive ions[8,32]. In ionization-based MS in general, the extent of charging required in some cases may make it complex to study the analyte in its "native" form, as it would be found in buffered solution[9], and may even induce unwanted dissociation. A clear limitation of ionization-based MS is the instruments' limit-of-detection, which is determined by the ionization source's ionization/sampling efficiency, as well as the efficiency of ion transmission and detection. For example, Electro Spray Ionization (ESI), the preferred source for use with macro-ions, has a combined ionization/sampling efficiency (ratio between the ionic current after the inlet capillary and the total spray current) well below 1%[14]. This is due in part to a rapidly expanding charged droplet/ion plume caused by Coulomb repulsion and to the shape of the electric field, which disperses the sample across a much larger area than that of the inlet to the mass spectrometer. Standard ion detectors also have a diminished efficiency at high masses due to their exponentially decreasing quantum efficiency[15]. During our experiments, only one particle out of a few million entering the TOF spectrometer was generally detected by the Micro Channel Plate detector (MCP). This is mainly due to the proportion of charged particles (25%) and to the secondary electron yield of the MCP (measured in the $10^{-6}$ range, as expected with this high mass range). The yield of the TOF-MCP pair can be improved by increasing the voltage, but not by much more than a factor of 50. On the other hand, one particle out of a hundred million particles was detected by our single NEMS detector despite its small cross-section (around 2 μm$^2$). We believe it is feasible to achieve ~10% detection efficiency through use of ultra-dense arrays of nanoscale resonators placed in a configuration covering the entire particle beam. Elsewhere we have demonstrated the requisite CMOS-compatible techniques for very large scale integration of NEMS[33].

The suitability of NEMS-MS for analysis of both biological and metallic particles in the MDa to the GDa range has already been shown[22]. In this study, we demonstrate that NEMS-MS directly measures mass independently of charge. This clarifies the spectrum obtained, simplifying analysis and facilitating identification of the sample. We demonstrate directly, for the first time, that NEMS-MS is compatible with analysis of *neutral* particles. This result makes development of weakly-ionizing or non-ionizing methods for the transfer of particles into the gas phase highly relevant. Micro-dispensing could be particularly suitable for high-mass entities. For example, Surface Acoustic Wave Nebulization is an ionization technique which has been shown to be even softer than ESI[34]. In addition, particles have almost no velocity as they reach the mass spectrometer inlet, and can thus be sampled with optimal efficiency. Combining nebulizers and ultra-dense arrays of NEMS offers the potential of new analytical capabilities and limits-of-detection several orders of magnitude lower than today's technology. These features could help elucidate unsolved questions in various fields, such as the study of new materials[8], large dendrimers in analytical chemistry[35,36], or easily-dissociated, poorly soluble membrane proteins[37] in structural biology. A similar combination could also be used to monitor nanoparticle levels in the environment and residual diseases in real-time.

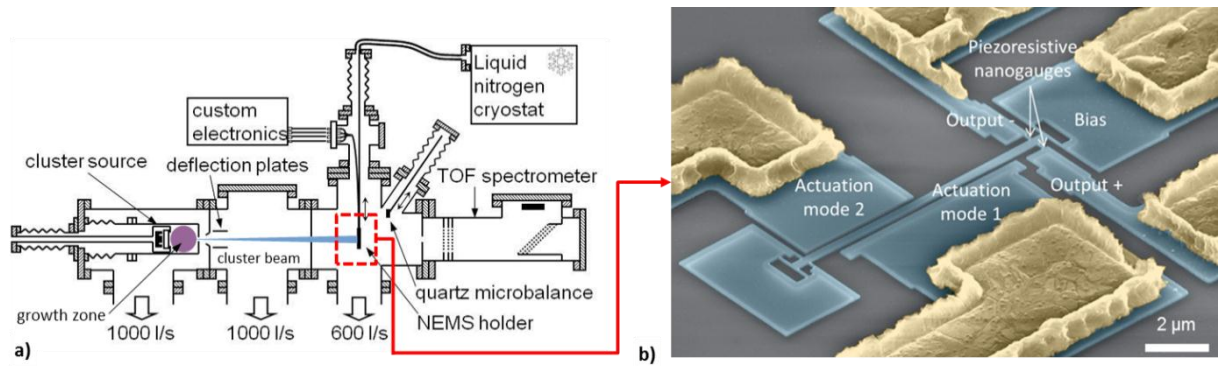

**Figure 1: Hybrid setup for TOF-MS and NEMS-MS of nanoparticles.** a) Diagram showing the full setup, with, from left to right: the cluster source, an intermediate chamber comprising deflection plates, the deposition chamber and an in-line TOF mass spectrometer. Both NEMS holder and QCM are retractable, making sequential NEMS-MS, TOF-MS and QCM measurements possible in identical operating conditions. b) **Colorized SEM image of a typical doubly clamped in-plane resonator used in this study.** The beams were designed to resonate around 25 MHz for mode 1 and 65 MHz for mode 2. Typical dimensions for the resonant beam were: 160 nm (thickness), 300 nm (width) and 5 to 10 µm (length). Electrostatic actuation was used, and in-plane motion transduction was performed using piezoresistive nanogauges in a bridge configuration to allow background cancellation. Gate electrodes were specifically patterned for mode 1 and mode 2 actuation. A bias voltage at $\omega-\Delta\omega$ was applied at the center of the nanogauge bridge and the tension/compression in the nanogauges at $\omega$ was used to mix the differential output down at $\Delta\omega$, a few 10s of kHz (see also Supplementary Methods). Manufacturing details and electrical operation are described elsewhere[31].

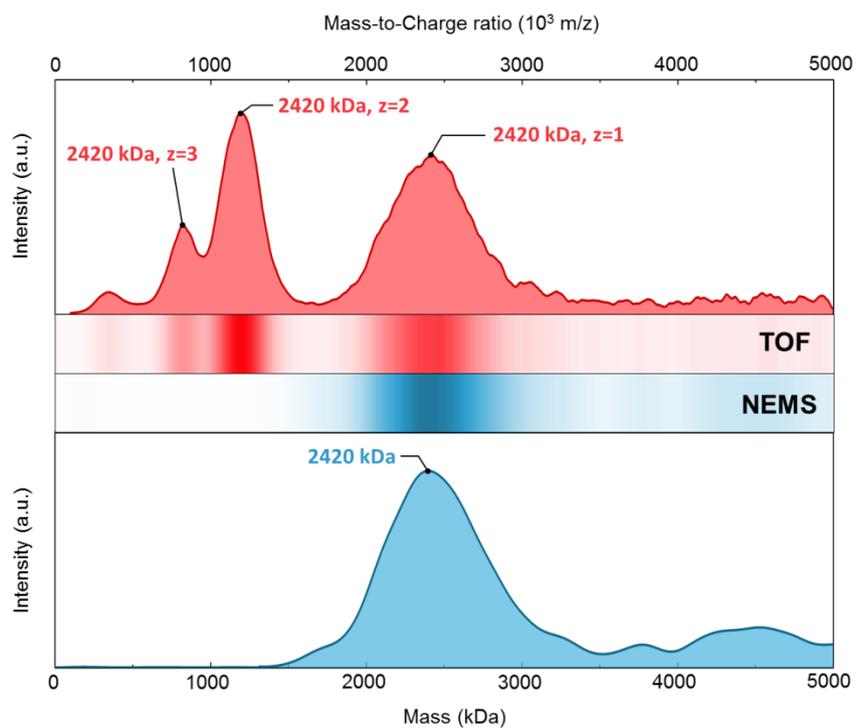

**Figure 2: NEMS-MS greatly clarifies the spectrum.** Spectra for a nanocluster population with a mean mass of around 2420 kDa acquired by both TOF and NEMS-MS. The TOF data was smoothed by applying a Savitsky-Golay filter. While multiply-charged particles generate several peaks in the TOF-MS spectrum, the NEMS directly weighs the particle masses, yielding a single peak.

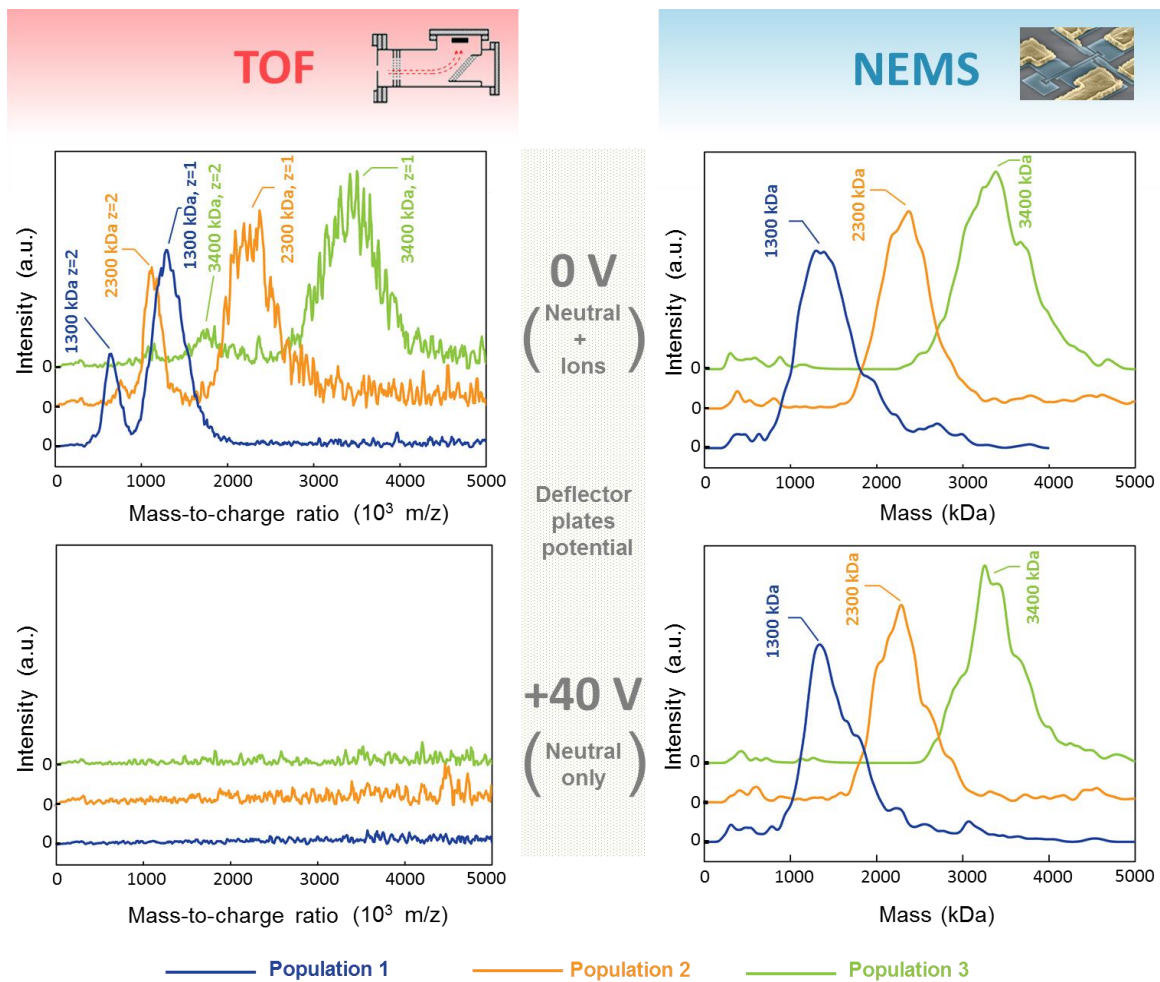

**Figure 3: Demonstration of neutral particle Mass Spectrometry.** Results of TOF-MS and NEMS-MS analysis of populations with and without charged particles. Charged particles can be removed by tuning the voltage on the deflection plates in the intermediate chamber: when this potential is set to 0 V (upper graphs), both ionized and neutral particles flow toward the deposition chamber, while with a voltage of 40 V (lower graphs) only neutral particles enter the deposition chamber. Three different nanoparticle populations were produced and measured using different experimental settings. These populations had mean masses of around 1300 kDa, 2300 kDa and 3400 kDa. The TOF displays singly, doubly and even triply charged particles when ions are not deflected. However, just like any ion-based MS instrument, the TOF is unable to produce spectra for the populations composed solely of neutral particles. NEMS-MS successfully detects the particles regardless of their charge states and delivers identical spectra for all configurations.

**Acknowledgment**

The authors acknowledge partial support from the LETI Carnot Institute NEMS-MS project. They also thank Carine Marcoux for her support with the device fabrication, A.-K. Stark for technical discussions as well as Prof. Jean-Claude Tabet for his advice on the writing of the manuscript. S.H. would like to thank Prof. Dan Frost for carefully proof-reading the text.

**Author contribution**

E.S., A.B. and R.M. performed the experiments. C.D. fabricated the devices. L.D., and S.H. conceived the idea, supervised the investigation and discussed the results and their impact with C.M.. E.S., A.B., T.A., C.M. and S.H. wrote the manuscript. All authors commented on the manuscript.

**Competing financial interests:** the authors declare no competing financial interests.